\documentclass[reprint,
superscriptaddress,
nobibnotes,
amsmath,amssymb,
aps,
longbibliography
]{revtex4-2}

\usepackage{dcolumn}
\usepackage{amssymb}
\usepackage{physics}
\usepackage{bm}
\usepackage{graphicx}
\usepackage[colorlinks,allcolors=blue]{hyperref}

\def\br{\bm{r}} % bold r 
\def\bk{\bm{k}} % bold k 
\def\bkp{\bm{k'}} % bold k'
\def\bq{\bm{q}} % bold q
\def\bG{\bm{G}} % bold G
\def\bv{\bm{v}} % bold v
\def\bF{\bm{F}} % bold F

\begin{document}
%\title{\textit{Ab initio} charged-defect-limited carrier mobility combining both electron-defect and electron-phonon interactions}
% : combining defect and phonon scattering in

\title{First-principles ionized-impurity scattering and charge transport in doped materials}
%transport calculations combining electron interactions \\with charged defects and phonons}

%\title{\textit{Ab initio} ionized impurity scattering and its inclusion in first-principles electrical transport calculations}

\author{I-Te Lu}
\author{Jin-Jian Zhou}
\author{Jinsoo Park}
\author{Marco Bernardi}
%\email{bmarco@caltech.edu}
\affiliation{Department of Applied Physics and Materials Science, California Institute of Technology, Pasadena, California 91125}

%\date{\today}

%%%%%%%%%
% Abstract
%%%%%%%%%
\begin{abstract}
% dominates low-temperature
Scattering of carriers with ionized impurities governs charge transport in doped semiconductors. %, oxides and other electronic materials.
% atomistic detail in real materials
\mbox{However,} electron interactions with ionized impurities cannot be fully described with quantitative first-principles calculations, so their understanding relies primarily on simplified models. 
%using first-principles approaches. or other charged defects 
% heuristic
Here we show an \textit{ab initio} approach to compute the interactions between electrons and ionized impurities or other charged defects. It includes the short- and long-range electron-defect ($e$-d) interactions on equal footing, and allows for efficient interpolation of the $e$-d matrix elements.
We combine the $e$-d and electron-phonon interactions in the Boltzmann transport equation to compute the carrier mobilities in doped silicon over a wide range of temperature and doping concentrations, spanning seamlessly the defect- and phonon-limited transport regimes. 
%
%No assumptions are made on the atomic structure of the defects, the electronic wave functions, and the electronic band structure of the material.
%
The individual contributions of the defect- and phonon-scattering mechanisms to the carrier relaxation times and mean-free paths are analyzed. 
%in the iterative solution of the linearized BTE. 
%that is consistent with the experimental data from the literature.
%
Our method provides a powerful tool to study electronic interactions in doped materials. It broadens the scope of first-principles transport calculations, enabling studies of a wide range of doped semiconductors and oxides with application to electronics, energy and quantum technologies.
\end{abstract}

\maketitle

%%%%%%%%%
% Introduction
%%%%%%%%%
%\section{Introduction}\label{sec:introduction}
%\vspace{-20pt}
%% 
%% broad introduction
%% describing these electronic processes 
Scattering of carriers with charged defects is treated primarily using heuristic models. An important example are calculations of ionized-impurity scattering in doped semiconductors and the resulting defect-limited transport properties. 
Widely adopted models neglect the detailed atomic structure of the defects, employ approximate defect-scattering potentials, and rely on simplified electronic band structures $-$ usually, a single isotropic band parametrized by the effective mass~\cite{chattopadhyay_electron_1981, Ridley, Lundstrom}.
% adopted and practical 
First-principles calculations of electron-defect ($e$-d) interactions for ionized impurities and other charged defects remain an open challenge. Their development would enable quantitative studies of transport in doped materials with complex atomic and electronic structures, providing microscopic insight into the effects of dopants in a wide range of technological materials and devices.
\\ 
\indent
%
%krishnaswamy_first-principles_2017, cao_dominant_2018, dsouza_electron-phonon_2020}. 
% ponce_towards_2018, ponce_structural_2020,
% first-principles calculations of ionized-impurity scattering and its impact on the carrier mobilit
Recent work has highlighted the computational cost and difficult workflows for obtaining $e$-d interactions in the framework on density functional theory (DFT)~\cite{restrepo_first-principles_2009, lu_efficient_2019, lu_ab_2020}, particularly when using fine Brillouin zone (BZ) grids required for transport calculations~\cite{bernardi_first-principles_2016, zhou_ab_2016, lu_efficient_2019, lu_ab_2020}. Various approximations have been employed to obviate these bottlenecks, such as treating the charged-defect perturbation as a Yukawa potential~\cite{qiu_first-principles_2015, Ganose2021} or the Bloch electronic states as plane waves~\cite{chaves_boosting_2020}. 
% simplifying approximations
Recent work by us~\cite{lu_efficient_2019, lu_ab_2020} and Kaasbjerg et al.~\cite{kaasbjerg_symmetry-forbidden_2017} has developed a fully \textit{ab initio} framework, based on DFT, to compute \textit{e}-d interactions without simplifying approximations, so far focusing on charge-neutral defects~\cite{kaasbjerg_symmetry-forbidden_2017, lu_efficient_2019, lu_ab_2020}.
Despite these advances, an efficient \textit{ab initio} method to compute \textit{e}-d interactions for charged defects and study their impact on charge transport is still missing. Ideally, such an approach would capture the atomic details of the defect perturbation potential, use electronic Bloch wave functions, and employ the full band structure of the material.
%take into account the band structure anisotropy and band multiplicity. 
%
%%% HERE WE SHOW
%
\\
\indent
Here, we show a first-principles method to efficiently compute the \textit{e}-d interactions for ionized impurities and other charged defects. Our approach is based on plane-wave DFT calculations and satisfies all the ideal characteristics listed above. 
% of the $e$-d perturbation 
To obtain the $e$-d matrix elements, we develop approaches for supercell potential-alignment and removal of spurious image contributions. Both the short- and long-range perturbations induced by the charged defects are included, without using any empirical or tuning parameters.
% on fine BZ grids using Wannier function (WF) interpolation, and
%To demonstrate the capabilities of our method, 
We combine the \textit{e}-d and electron-phonon (\textit{e}-ph) interactions in the Boltzmann transport equation (BTE) to compute the carrier mobility in a doped material with an accurate account of ionized-impurity scattering. 
We apply our method to  silicon doped with phosphorous (P) or boron (B); we compute and analyze state-dependent relaxation times (RTs) for ionized-impurity scattering, and predict the doping and temperature dependence of the electron and hole mobilities in quantitative agreement with experiment.  
Our treatment of electron-charged defect interactions complements recent efforts to develop quantitative tools to study charge transport in real materials~\cite{zhou_ab_2016, Jhalani-quad, Perturbo, lu_ab_2020, Desai, ponce_towards_2018, Brunin}. 
%
%%%%%%%%%%
%% Methodology
%%%%%%%%%%
%
%%%%%%%%%%%%%%%%%%%%%%%%%%%%%
% Electron-defect interaction for charged defects
%%%%%%%%%%%%%%%%%%%%%%%%%%%%%
%
% Define the e-d matrix elements 
%
\\
\indent 
The $e$-d matrix element coupling a Bloch state $\ket{n\bk}$, with band index $n$ and crystal momentum $\bk$, to another state $\ket{m\bkp}$ due to the perturbation potential from a charged defect, $\Delta V_{e-\rm{d}}^{\rm{cd}}$, is defined as
\begin{equation}\label{eq:def-edmatcd}
M_{mn}^{\rm{cd}}(\bkp,\bk)=\mel{m\bkp}{\Delta V_{e-\rm{d}}^{\rm{cd}}}{n\bk},
\end{equation}
where the Bloch states are for the pristine system without any defect~\cite{lu_efficient_2019}. 
% $\Delta V^{\rm{cd}}$
The $e$-d defect perturbation potential is obtained from DFT calculations as the difference between the Kohn-Sham (KS) potential of the pristine system and that of the same system containing a defect~\cite{lu_efficient_2019}.
%
%The Hamiltonian is chosen as the Kohn-Sham (KS) Hamiltonain $H_{\rm{KS}}$, which can be routinely obtained within density functional theory (DFT). 
%
We use two supercells with the same size to simulate the pristine and defect-containing systems. 
For charged defects, the difference between the KS Hamiltonians of these two supercells, $\Delta V_{\rm{KS}} = H_{\rm{KS}}^{(\rm{d})}-H_{\rm{KS}}^{(\rm{p})}$, is not equal to the $e$-d perturbation, and requires a correction that is not needed for charge-neutral defects~\cite{lu_efficient_2019}. 
%
% Deal with charged defects in supercells 
%
\\
\indent
To treat charged defects in plane-wave DFT, the widely-used Jellium model~\cite{van_de_walle_first-principles_2004} removes the divergence due to the charge introduced in the supercell by setting to zero the $\bG_{\rm{sup}}\!\!=\!0$ component of the Hartree potential ($\bG_{\rm{sup}}$ are reciprocal lattice vectors of the supercell).
%; for other methods that are beyond the jellium model, please see Sec.~\ref{sec:discussion} for discussion. 
%
%Here we add (or remove) extra electrons into (from) the supercell containing the charged defect and set the $\bG_{\rm{sup}}=0$ component of the KS potential $-$ $\bG_{\rm{sup}}$ is the reciprocal lattice vectors of the supercell $-$ to zero to remove the divergence due to the extra charges.
% ~\cite{ohara_formation_2019}
\mbox{Although} the divergence has been removed, the supercell is, in a sense, still charged $-$ since a uniform electron density is added to the system, the long-range tail of the Coulomb potential generated by the charged defect is still present~\cite{ohara_formation_2019}. 
As a result, the charged-defect perturbation potential does not decay to zero inside the supercell, and therefore is contaminated by the periodic images of the supercell. %calculation.
%

%
% Remove the long-range contribution 
%
To remove these spurious contributions, we introduce a supercell-periodic screened-Coulomb potential, $V^{\rm{scr}}_{\rm{sup}}$, generated from a point charge $+Ze$ (where $e$ is the electron charge) in the supercell, which simulates the charge state of the defect: %_{\rm{sup}}
\begin{equation}
V^{\rm{scr}}_{\rm{sup}}(\br) = \frac{1}{\Omega_{\rm{sup}}}\sum_{\bG_{\rm{sup}}\neq 0}\widetilde{V}^{\rm{scr}}(\bG_{\rm{sup}})e^{i\bG_{\rm{sup}}\cdot\br},
\end{equation}
where $\Omega_{\rm{sup}}$ is the volume of the supercell. %and $\bG_{\rm{sup}}$ are supercell reciprocal lattice vectors.
The Fourier coefficient at wave vector $\bq$ of this screened-Coulomb potential is defined as 
\begin{equation}
\widetilde{V}^{\rm{scr}}(\bq) = \frac{-Z e^{2}}{\epsilon_{0}\epsilon(q)\, q^{2}},
\end{equation}
where $\epsilon_{0}$ is the vacuum permittivity and $\epsilon(q)$ is the dielectric function due to the screening from the valence electrons, described here using an isotropic homogeneous electron model. %to describe the screening from the valence electrons. 
%
%One can also use the dielectric tensor instead of the dielectric function to take the anisotropic effect into account~\cite{rurali_theory_2009, murphy_anisotropic_2013}.
% remove the spurious Coulomb potential from the images
The spurious image-cell contributions are then removed by subtracting $V^{\rm{src}}_{\rm{sup}}$ from the KS perturbation potential. 
An additional correction is needed to account for DFT potential-alignment errors due to the different potential references used in the pristine and defect-containing supercell calculations~\cite{lu_efficient_2019}. 
With these corrections, we compute the neutral part of the $e$-d perturbation potential as
% ~\cite{}. $V_{\rm{align}}$ $-$ which is needed because of
\begin{equation}
\Delta V_{e-\rm{d}}^{\rm{np}}=\Delta V_{\rm{KS}}-V^{\rm{scr}}_{\rm{sup}}-V_{\rm{align}},
\end{equation}
where the $V_{\rm{align}}$ term enforces the potential alignment.
%
% Add the long-range Coulomb potential of an isolated charged defect back 
%
\\
\indent
Finally, we obtain the charged-defect perturbation potential, $\Delta V_{e-\rm{d}}^{\rm{cd}}$, by adding the screened-Coulomb potential due to an \textit{isolated} point charge, $V^{\rm{scr}}_{\rm{ex}}(n_{\rm{ex}})$; the latter is screened by both the valence electrons and by any extrinsic free carriers (with concentration $n_{\rm{ex}}$) resulting from ionized impurities:
\begin{equation}\label{eq:cdpot}
\Delta V_{e-\rm{d}}^{\rm{cd}} = \Delta V_{e-\rm{d}}^{\rm{np}} + V^{\rm{scr}}_{\rm{ex}}(n_{\rm{ex}}),
\end{equation}
with the Fourier coefficient of the screened-Coulomb potential of the isolated defect computed as~\cite{resta_dielectric_1979}
\begin{equation}\label{eq:coeff-vscr}
\widetilde{V}^{\rm{scr}}_{\rm{ex}}(\bq, n_{\rm{ex}}) = \frac{-Ze^{2}}{\epsilon_{0}[\epsilon(q)q^{2}+\epsilon(0)q_{\rm{scr}}^{2}]}.
\end{equation}
Above, $\epsilon(0)$ is the dielectric constant, and the inverse screening length $q_{\rm{scr}}$ due to the free carriers is~\cite{ashcroft1976solid}
\begin{equation}
q_{\rm{scr}}=\left[\frac{e^{2}}{\epsilon_{0}\epsilon(0)}\int_{E_{\rm{c}}}^{\infty} dE\, g(E)\left(-\frac{\partial f^{0}(E)}{\partial E}\right)\right]^{1/2},
\end{equation}
where $g(E)$ is the density of states (DOS) at electron energy $E$, $E_{\rm{c}}$ is the conduction band minimum, and $f^{0}(E)$ is the Fermi-Dirac distribution. 
This expression is for electrons, but a similar one holds for hole carriers~\cite{ashcroft1976solid}. 
%
%
% e-d matrix elements to compute 
%
\\
\indent
Substituting Eq.~(\ref{eq:cdpot}) into Eq.~(\ref{eq:def-edmatcd}) splits the $e$-d matrix element for a charged defect into two parts, 
\begin{equation}
M^{\rm{cd}}_{mn}(\bkp,\bk) = M^{\rm{np}}_{mn}(\bkp,\bk)+M^{\rm{scr}}_{mn}(\bkp,\bk).
\end{equation}
The first term, $M^{\rm{np}}_{mn}(\bkp,\bk) \!\!=\!\! \mel{m\bkp}{\Delta V_{e-\rm{d}}^{\rm{np}}}{n\bk}$, is a short-ranged  charge-neutral contribution, while the second term, $M^{\rm{scr}}_{mn}(\bkp,\bk) \!=\! \mel{m\bkp}{V^{\rm{scr}}_{\rm{ex}}(n_{\rm{ex}})}{n\bk}$, accounts for the long-range screened Coulomb interaction generated by the charged defect (with the spurious image contributions properly removed).
In practice, the $e$-d matrix elements are computed using the approach developed in Ref~\cite{lu_efficient_2019}, using wave functions from a \textit{primitive} unit cell (as opposed to the large supercell) to greatly reduce the computational cost.
The Wannier function (WF) interpolation scheme we recently developed for charge-neutral defects~\cite{lu_ab_2020} is then employed to interpolate the neutral part of the matrix elements, $M^{\rm{np}}_{mn}(\bkp,\bk)$, on ultra-fine BZ grids.
For the long-range part, $M^{\rm{scr}}_{mn}(\bkp,\bk)$, we develop a different interpolation approach inspired by recent work on $e$-ph interactions in polar materials~(for details, see the Supplemental Material~\cite{SM}).  
%
%% 
% Carrier mobility due to electron-phonon and electron-defect interaction
%% 
\\
\indent
With the \textit{ab initio} ionized-impurity scattering in hand, we combine $e$-ph and elastic $e$-d interactions in the BTE framework to compute the carrier mobility as a function of temperature and doping concentration~\cite{li_electrical_2015, bernardi_first-principles_2016}. 
% mobility $\mu_{\alpha\beta} $ can be obtained from the formula $\sigma_{\alpha\beta}/(en_{c})$, where $\alpha$ and $\beta$ are Cartesian directions and $n_{c}$ is the carrier concentration, and the 
The conductivity tensor $\sigma_{\alpha\beta}$ is computed using~\cite{Ziman,Mahan,Perturbo}
\begin{equation}
\label{eq:conduct}
\sigma_{\alpha\beta}=e^{2}\int dE \left(-\partial f^{0}/\partial E\right)\Sigma_{\alpha\beta}\left(E\right),
\end{equation}
where $\alpha$ and $\beta$ are Cartesian directions; the mobility is obtained using $\mu_{\alpha\beta} = \sigma_{\alpha\beta}/(en_{c})$, where $n_{c}$ is the carrier concentration. The transport distribution function $\Sigma_{\alpha\beta}\left(E\right)$ is defined as~\cite{Perturbo} 
\begin{equation}
\Sigma_{\alpha\beta}\left(E\right)=\frac{S}{N_{\bk}\Omega_{\rm{uc}}}\sum_{n\bk}\bv_{n\bk}^{\alpha}\bF_{n\bk}^{\beta}\delta\left(E-E_{n\bk}\right),
\end{equation}
where $S$ is the spin degeneracy and $N_{\bk}$ is the number of $\bk$-points in the BZ; $E_{n\bk}$ and $\bv_{n\bk}$ are electron band energies and velocities, respectively.
The vector $\bF_{n\bk}$ is proportional to the steady-state occupation change of each electronic state~\cite{Perturbo},
%
% F_{nk} vector 
% The vector $\bF_{n\bk}$ can
and is obtained by solving the linearized BTE for a weak electric field. Extending the approach developed in {\sc{Perturbo}}~\cite{Perturbo} to include both $e$-d and $e$-ph interactions, the BTE becomes:
\begin{equation}\label{eq:vectorF}
\begin{split}
\bF_{n\bk}\left(\Gamma_{n\bk}^{e-\rm{ph}}+\Gamma_{n\bk}^{e-\rm{d}}\right) 
& = \bv_{n\bk}+\frac{1}{N_{\bq}}\sum_{m,\nu\bq}\bF_{m\bk+\bq}W_{m\bk+\bq,n\bk}^{\nu\bq,\,e-\rm{ph}} \\ 
& +\frac{1}{N_{\bkp}}\sum_{m\bkp\neq n\bk}\bF_{m\bkp}W_{m\bkp,n\bk}^{e-\rm{d}},
\end{split}
\end{equation}
where $N_{\bq}$ and $N_{\bkp}$ are the numbers of $\bq$-points and $\bkp$-points in the BZ, respectively.
The $e$-ph scattering rate for each electronic state, $\Gamma_{n\bk}^{e-\rm{ph}}$, is computed using~\cite{Perturbo} 
\begin{equation}\label{eq:eph-scatrate}
\Gamma_{n\bk}^{e-\rm{ph}} = \frac{1}{N_{\bq}}\sum_{m,\nu\bq}W_{m\bk+\bq,n\bk}^{\nu\bq,\,e-\rm{ph}},
\end{equation}
where $W_{m\bk+\bq,n\bk}^{\nu\bq,\,e-\rm{ph}}$ is the scattering rate from state $\ket{n\bk}$ to $\ket{m\bk+\bq}$ due to the emission or absorption of a phonon with mode index $\nu$ and crystal momentum $\bq$~\cite{Perturbo}.
%
%\begin{equation}
%\begin{split}
%W_{m\bk+\bq,n\bk}^{\nu\bq,\,e-\rm{ph}} 
%& =\frac{2\pi}{\hbar}|g_{mn\nu}(\bk,\bq)|^{2}\\ 
%& \times[\delta(\varepsilon_{n\bk}-\hbar\omega_{\nu\bq}-\varepsilon_{m\bk+\bq})(1+N^{0}_{\nu\bq}-f_{m\bk+\bq}^{0})\\
%& + \delta(\varepsilon_{n\bk}+\hbar\omega_{\nu\bq}-\varepsilon_{m\bk+\bq})(N^{0}_{\nu\bq}+f_{m\bk+\bq}^{0})],
%\end{split}
%\end{equation}
%
%where $g_{mn\nu}(\bk,\bq)$ is the $e$-ph matrix element that couples the state $\ket{n\bk}$ with the state $\ket{m\bk+\bq}$ via the phonon mode $\ket{\nu\bq}$ with a phonon frequency $\omega_{\nu\bq}$, $N^{0}_{\nu\bq}$ is the Bose-Einstein distribution function for the phonon mode, and $\hbar$ is the reduced Planck constant.
%
The $e$-d scattering rate is defined analogously as 
\begin{equation}\label{eq:ed-scatrate}
\Gamma_{n\bk}^{e-\rm{d}} = \frac{1}{N_{\bkp}}\sum_{m\bkp\neq n\bk}W_{m\bkp,n\bk}^{e-\rm{d}},
\end{equation}
where $W_{m\bkp,n\bk}^{e-\rm{d}}$ is the elastic $e$-d scattering rate between two electronic states due to a charged defect~\cite{lu_efficient_2019}:
\begin{equation}\label{eq:ed-wprop}
W_{m\bkp,n\bk}^{e-\rm{d}}=\frac{2\pi}{\hbar}n_{\rm{at}}C_{\rm{imp}}\left|M_{mn}(\bkp,\bk)\right|^{2}\delta\left(E_{m\bkp}-E_{n\bk}\right).
\end{equation}
% the state $\ket{n\bk}$ is excluded in the summation to avoid the scattering into the initial state itself.
In this expression, $n_{\rm{at}}$ is the number of atoms in the primitive cell and the scattering rate is proportional to the impurity concentration $C_{\rm{imp}}$, defined as in Ref.~\cite{lu_efficient_2019} as the dimensionless ratio of the number of impurities to the total number of atoms in the crystal.
The $e$-ph and $e$-d relaxation times (RTs) for each electronic state are obtained as the inverse of the respective scattering rates,  $\tau_{n\bk}^{e-\rm{ph}}=(\Gamma_{n\bk}^{e-\rm{ph}})^{-1}$ 
and $\tau_{n\bk}^{e-\rm{d}}=(\Gamma_{n\bk}^{e-\rm{d}})^{-1}$. 
\\
\indent
The occupation change vector $\bF_{n\bk}$ in Eq.~(\ref{eq:vectorF}) can be obtained with an iterative approach (ITA) to solve the BTE~\cite{li_charge_2016, Perturbo}, or alternatively by using the relaxation-time approximation (RTA), where backscattering is neglected by setting to zero the two summations in the right-hand side of Eq.~(\ref{eq:vectorF}). 
For elastic $e$-d interactions, a widely used approach to \text{approximately} account for backscattering is to use the RTA with the $e$-d scattering rates multiplied by a cosine factor~\cite{Mahan}.
%  of ($1-\cos\theta_{m\bkp,n\bk}$)
%to take the back scattering into account, and the scattering rate 
These so-called \textit{transport}~$e$-d scattering rates are defined as~\cite{Mahan}:
\begin{equation}
\Gamma_{n\bk}^{e-\rm{d},\,\rm{tr}} =  \frac{1}{N_{\bkp}}\sum_{m\bkp\neq n\bk}W_{m\bkp,n\bk}^{e-\rm{d}}(1-\cos\theta_{m\bkp,n\bk}),
\end{equation}
where $\theta_{m\bkp,n\bk}$ is the angle between the band velocities $\bv_{m\bkp}$ and $\bv_{n\bk}$. Using these transport $e$-d scattering rates in the left-hand side of Eq.~(\ref{eq:vectorF}), while leaving out the backscattering terms in the right-hand side, leads to the transport relaxation-time approximation (tr-RTA). 
%The associated transport RT is  $\tau_{n\bk}=(\Gamma_{n\bk}^{e-\rm{d},\, t})^{-1}$; when used in the left-hand side of Eq.~\ref{eq:b}
%
%We use the `momentum' scattering rate to point out the subtle difference from the scattering rate.  
%
\\
\indent
In this work, we first obtain the occupation changes $\bF_{n\bk}$ by solving the BTE in one of the three flavors described above (ITA, RTA or tr-RTA). We then compute the conductivity using Eq.~(\ref{eq:conduct}), and from it obtain the carrier mobility $\mu = \sigma/(en_{c})$. 
Our approach allows us to include scattering from both the $e$-d interactions due to charged defects (here, ionized impurities) and the $e$-ph interactions.    
We can also obtain the mobility limited by only one of the $e$-ph or $e$-d interactions, by leaving out, respectively, the defect or phonon scattering terms in the BTE.
The formalism discussed above has been implemented in our open-source code, {\sc{Perturbo}}~\cite{Perturbo}. %, which is also used to compute the $e$-ph and $e$-d matrix elements.
\\
\indent
%
%
%
%
%
%%%%%%%%%%
%% Results
%%%%%%%%%%
%
%
%%%%%%%%%%%%%%%%%
%%%%%%  Figure 1 %%%%%%
%%%%%%%%%%%%%%%%%
\begin{figure}[t]
\centering
\includegraphics[width=\linewidth]{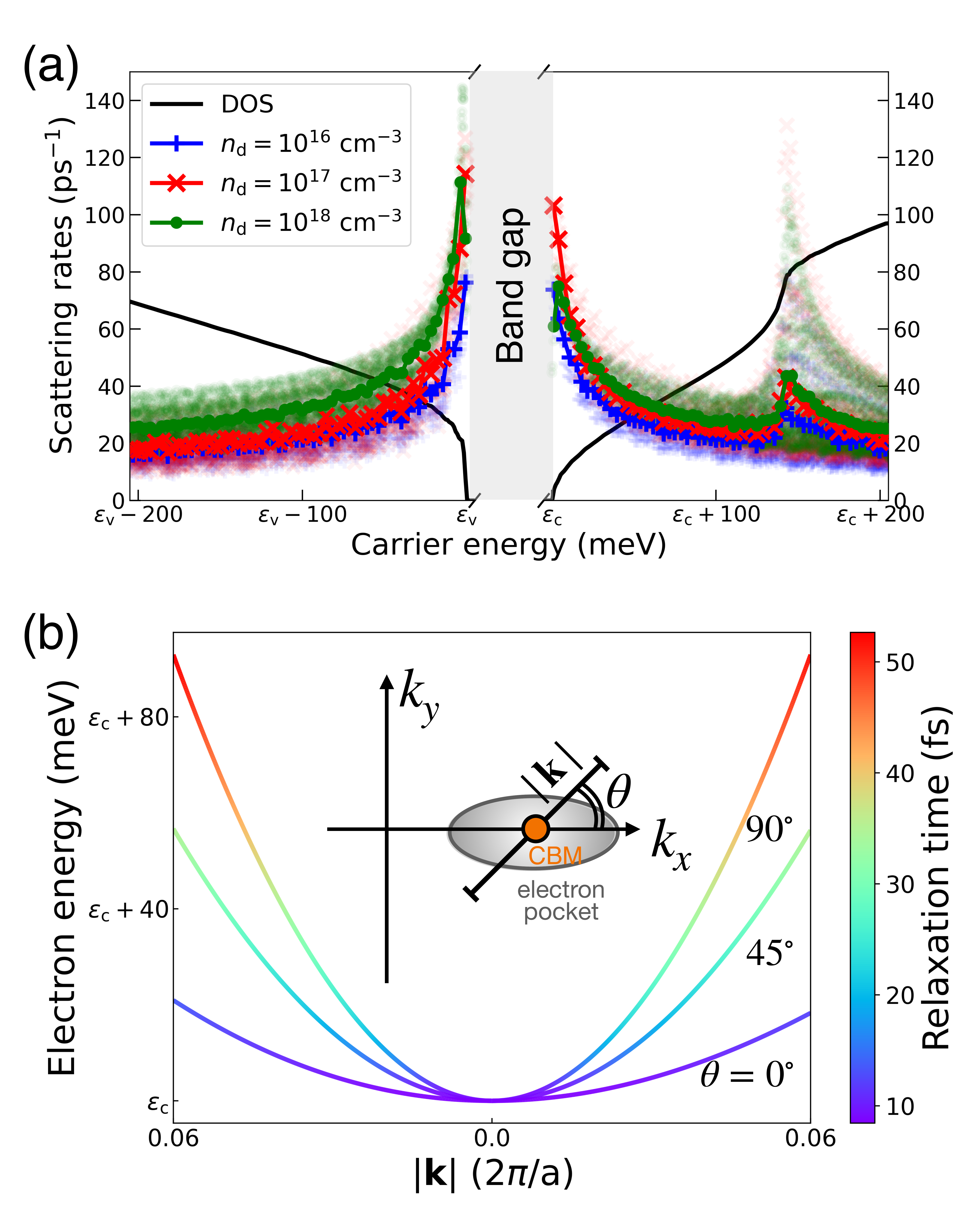}
\caption{
\textit{Ab initio} $e$-d scattering rates and RTs in doped Si at $300$~K.
(a) Scattering rates and DOS as a function of energy for various doping concentrations $n_{\rm{d}}$. 
%Results for electron and hole carriers are obtained, respectively, from calculations on the majority carriers in P-doped and B-doped Si. 
%
The shaded data points are raw values while the solid lines are scattering rates averaged over a small energy window for better visualization. The calculations include multiple bands $-$ the three highest valence bands for holes and the two lowest conduction bands for electrons. 
(b) Electron RTs along different paths passing through the CBM, obtained for $n_d \!=\! 10^{17}$~cm$^{-3}$. 
}
\label{fig:edtau-1st}
\end{figure}
We apply our approach to Si doped with phosphorous (P) or boron (B), respectively to study electron or hole carriers; 
%The results discussed below for electron and hole carriers are obtained, respectively, from calculations on P-doped and B-doped Si (for each case, we study the majority carriers). 
for now, we assume that the dopants are fully ionized. 
We carry out plane-wave DFT calculations~\cite{perdew_generalized_1996} with norm-conserving pseudopotentials~\cite{van_setten_pseudodojo_2018} using the {\sc{Quantum Espresso}} code~\cite{giannozzi_quantum_2009} (numerical details are provided below~\cite{comp-detail}). 
Figure~\ref{fig:edtau-1st}(a) shows our computed $e$-d scattering rates for electron and hole majority carriers in doped Si at $300$ K, for doping concentrations of $10^{16}$$-$$10^{18}$ cm$^{-3}$.
The $e$-d scattering rates decrease for increasing carrier energy, referenced to the valence band maximum (VBM) for holes and conduction band minimum (CBM) for electrons. This trend is opposite to neutral-defect scattering~\cite{lu_efficient_2019}, where the scattering rates increase with carrier energy due to the greater electronic DOS, which measures the number of accessible final states for elastic $e$-d scattering. 
Our results show that this phase-space argument does not hold for charged defects, where the momentum dependence of the $e$-d matrix elements plays a dominant role $-$ since the matrix elements increase rapidly for decreasing transferred momenta $|\bkp - \bk|$, 
the $e$-d scattering rates are greater near the valence and conduction band edges, despite the small DOS values.
For the same reason, we find a peak in the electron scattering rates $\sim$$\,$$150$~meV above the CBM in correspondence with a band crossing. Note also how increasing doping levels lead to stronger $e$-d scattering (and thus greater scattering rates), a relevant sanity check for our calculations.
\\
\indent
Our approach can capture the momentum dependence of the scattering rates. At each given carrier energy, the scattering rates exhibit a range of values due to the anisotropic character of the band structure and scattering processes.
To highlight this point, Fig.~\ref{fig:edtau-1st}(b) shows the state-dependent RTs for electronic states in the lowest conduction band along three different paths passing through the CBM. 
These paths, chosen to lie in the $k_{x}$-$k_{y}$ plane, are specified by an angle $\theta$ measured from the $k_{x}$ axis (i.e., the $\Gamma-$X direction in the BZ), so that $\theta=0^{\circ}$ is the longitudinal, and $\theta=90^{\circ}$ the transverse valley direction~\cite{Lundstrom}.
For a fixed carrier energy $-$ here, $20$ meV above the CBM $-$ the RTs for $\theta=0^{\circ}$, $45^{\circ}$, and $90^{\circ}$ are $\sim$10, 20, and 30 fs respectively, thus demonstrating that electronic states with the same energy can have a broad distribution of RTs. 
Our ability to treat $e$-d scattering for anisotropic, multi-valley band structures is key for studies of defect-limited transport in complex materials. 
\\
\indent
%
%
%
%
%%%%%%%%%%%%%%%%%
%%%%%%  Figure 2 %%%%%%
%%%%%%%%%%%%%%%%%
%
\begin{figure}[t]
\centering
\includegraphics[width=\linewidth]{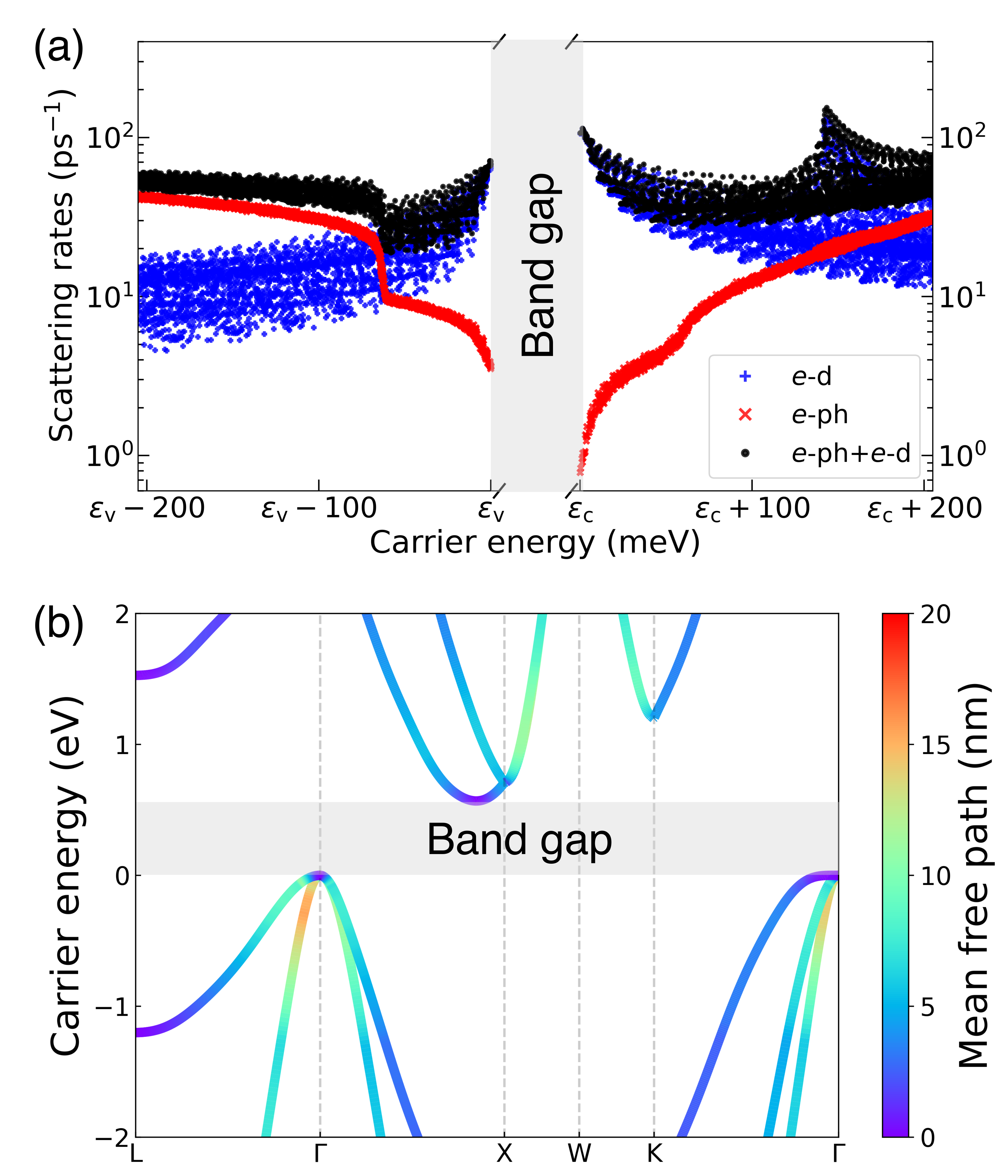}
\caption{
%Total scattering rates and the corresponding mean free path of the majority carrier due to $e$-ph and $e$-d scattering in P-doped and B-doped Si with a doping concentration of $10^{17}$ cm$^{-3}$ at $300$ K. 
%
(a) Scattering rates due to each of the $e$-ph and $e$-d ionized-impurity interactions, and their combination obtained via Matthiessen's rule.
(b) The corresponding mean free paths due to both $e$-ph and $e$-d scattering mapped onto the band structure. All results are calculated for Si at 300~K for a doping concentration of $10^{17}$ cm$^{-3}$. 
}
\label{fig:ephd-scat-mfp}
\end{figure}
Next, we examine the individual and combined effects of the $e$-ph and $e$-d interactions in a doped material [see Fig.~\ref{fig:ephd-scat-mfp}(a)]. 
%shows calculations of the $e$-ph scattering rate (including all phonon modes) and . 
%shows the scattering rates for the majority carriers due to $e$-ph and $e$-d interaction in P-doped and B-doped Si with a doping concentration of $10^{17}$ cm$^{-3}$ at $300$ K. %(see Sec.~\ref{sec:details} for more computational details.) 
%
%The \textit{ab initio} $e$-ph scattering rates here include all the phonon modes. 
%
Different from the ionized-impurity case, the $e$-ph scattering rates increase with carrier energy due to the dominant role of the scattering phase-space.
As a consequence, $e$-d scattering from ionized impurities dominates at low energy and $e$-ph scattering at higher carrier energies. 
%
%A similar feature for electrons has also been observed in a study~\cite{qiu_first-principles_2015}, which combines $e$-d `momentum' scattering rates and $e$-ph scattering rates but for only the lowest conduction band.
%
This result shows that different scattering mechanisms can govern electron dynamics in different electron energy windows, a valuable physical insight for transport and device physics. 
\\
\indent
As an example, a key material property in nano- and micro-electronic devices is the electron mean free path (MFP), namely the distance traveled by the carriers between scattering events~\cite{datta_electronic_1997, kasap_springer_2017, boer_semiconductor_2018}. 
% of an $e$-ph or $e$-d for $10^{17}$ cm$^{-3}$ doping are given
We compute state-dependent electron MFPs~\cite{bernardi-si}, $L_{n\bk}=\tau_{n\bk} |\bv_{n\bk}|$, using RTs that include both the $e$-ph and $e$-d interactions (via Matthiessen's rule~\cite{Ziman}). 
The computed MFPs for doped Si at 300~K, shown in Fig.~\ref{fig:ephd-scat-mfp}(b),
are minimal near the band edges due to the strong ionized-impurity scattering and small band velocities, and increase non-monotonically within $1$ eV of the band edges. 
%MARCO: (neglecteddue to the increasing $e$-ph scattering rates.) Great e-ph scattering rates alone would decrease the MFP
% 
The longest MFPs, of order 10$-$20~nm, are achieved $200$ meV above the CBM for electrons and $300$ meV below the VBM for holes. The ability to find energy windows with optimal MFPs can be leveraged to design hot-carrier devices for energy and sensing applications~\cite{hc-1,hc-2}.
\\
\indent
%
%%%%%%%%%%%%%%%%%
%%%%%%  Figure 3 %%%%%%
%%%%%%%%%%%%%%%%%
%
\begin{figure}[t]
\centering
\includegraphics[width=\linewidth]{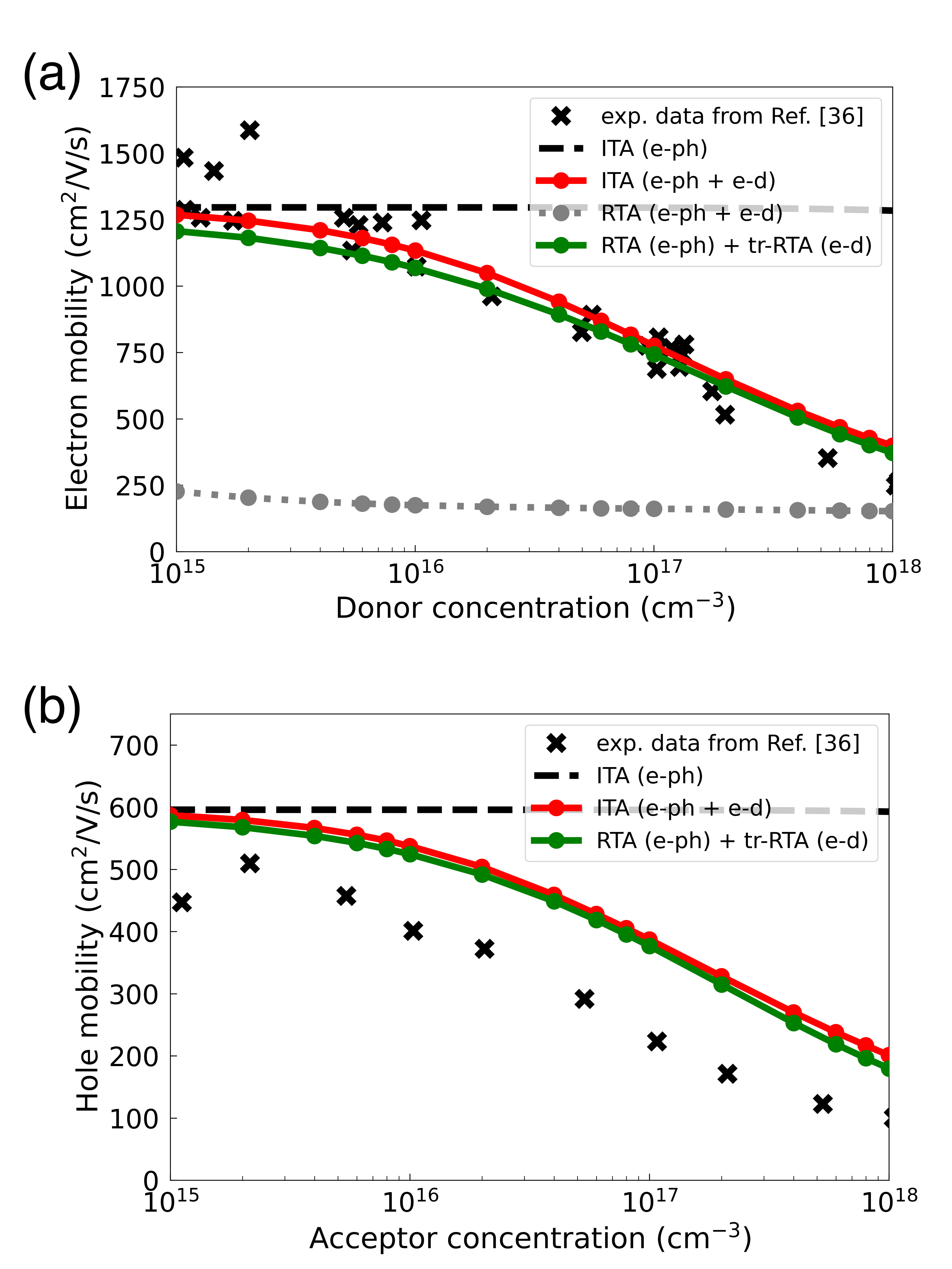}
\caption{
Carrier mobility in doped Si at 300~K. 
(a) Electron mobility in P-doped Si as a function of donor concentration.
(b) Hole mobility in B-doped Si as a function of acceptor concentration. 
}
\label{fig:mobility}
\end{figure}
Predicting the mobility in doped semiconductors is critical to designing electronic and energy devices~\cite{Lundstrom, kasap_springer_2017}. Most first-principles calculations have so far focused on the phonon-limited mobility~\cite{zhou_ab_2016, Brunin}, in some cases adding ionized-impurity scattering with simple models~\cite{ponce_towards_2018}. 
% effective-mass
Here our goal is to capture both phonon and impurity scattering in a fully first-principles quantitative framework~\footnote{A first step in this direction, using methods less advanced than those developed here, was taken in early work by Restrepo et al.~\cite{restrepo_first-principles_2009}.}.
% , restrepo_first_2014
\\
\indent
We first focus on the dependence of the electron and hole mobilities on doping concentration. 
Figure~\ref{fig:mobility}(a) compares the computed electron mobility with experiments~\cite{jacoboni_review_1977} for P-doped Si at $300$~K. 
The phonon-limited mobility, computed using the ITA without $e$-d interactions, is independent of donor concentration and overestimates the experimental mobility values at all doping concentrations greater than 10$^{15}$~cm$^{-3}$.  
Including both $e$-ph and $e$-d ionized-impurity scattering within the ITA allows us to predict the experimental data with a high accuracy up to $n_d \!=\! 10^{18}$ cm$^{-3}$. Greater doping concentrations that modify the band structure~\cite{altermatt_simulation_2006} are not studied here.   
We also find that using the RTA for $e$-ph plus tr-RTA for $e$-d scattering provides electron mobilities close to the full-ITA solution (with $e$-ph plus $e$-d interactions) for a greatly reduced computational cost. 
Conversely, the RTA for both $e$-ph and $e$-d scattering (i.e., without the cosine factors in the $e$-d scattering) significantly underestimates the mobility. 
\\
\indent
We obtain analogous results for hole carriers in B-doped Si at $300$ K [Fig.~\ref{fig:mobility}(b)]. Comparing the computed hole mobility with experimental data~\cite{jacoboni_review_1977} again shows that full ITA calculations 
%including both $e$-ph and ionized impurity $e$-d scattering 
can correctly predict the dependence of the mobility on doping, providing results in agreement with experiment. 
Similar results are obtained with $e$-ph RTA plus $e$-d tr-RTA calculations. For hole carriers, we find a greater discrepancy with experiment (a factor of 1.2$-$2x) than for electrons. Improving the effective masses (for example, using GW or experimental values~\cite{ponce_towards_2018}) and accounting for spin-orbit coupling would refine the results. 
\\
\indent
%
%%%%%%%%%%%%%%%%%
%%%%%%  Figure 4 %%%%%%
%%%%%%%%%%%%%%%%%
%
\begin{figure}[b]
\centering
\includegraphics[width=\linewidth]{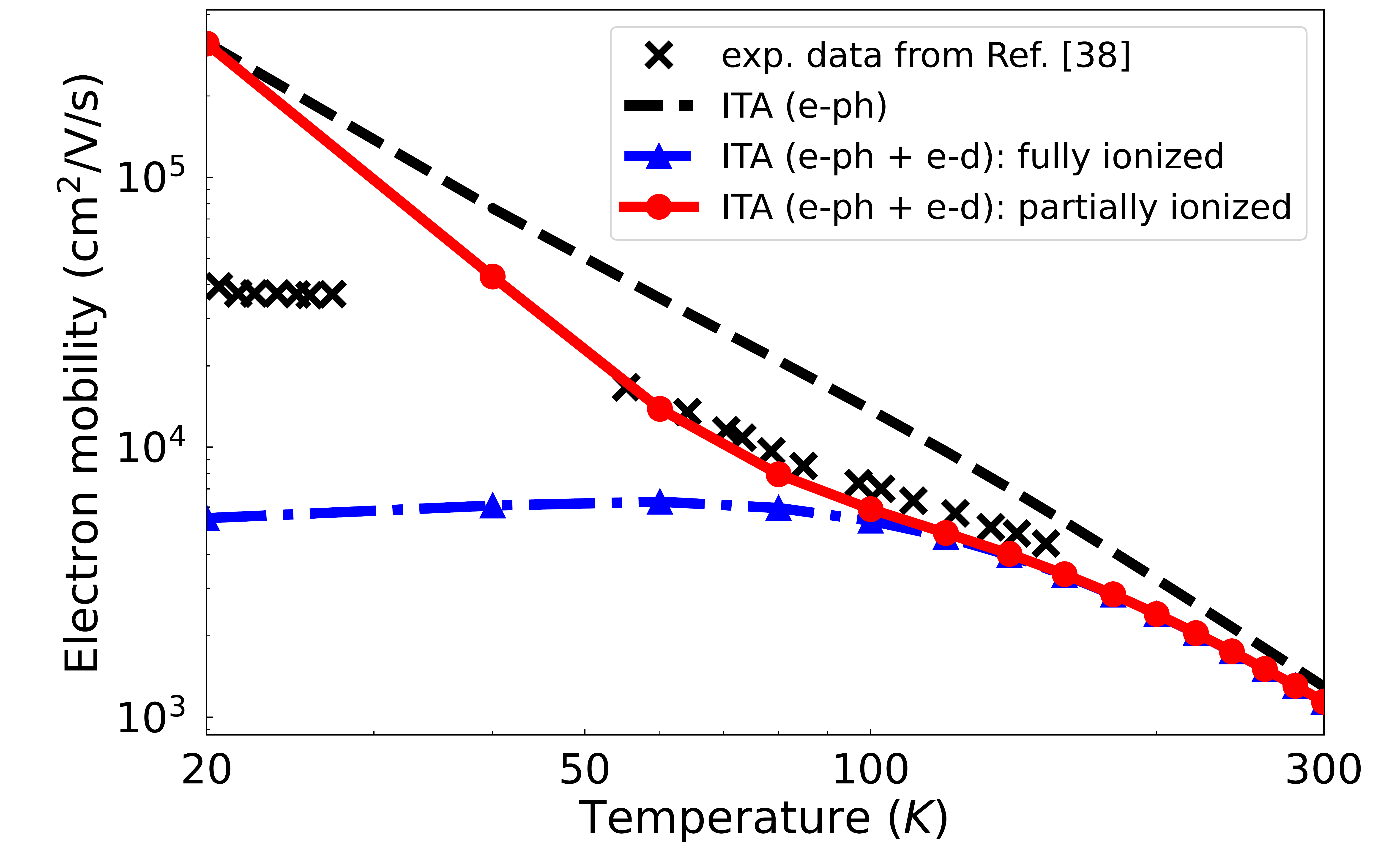}
\caption{
Electron mobility as a function of temperature in P-doped Si with a doping concentration $n_d = 9.5\times10^{15}$ cm$^{-3}$.
}
\label{fig:mob-temper}
\end{figure}
Finally, we analyze the temperature dependence of the mobility for a fixed doping concentration. 
%For example, many electronic devices operate at low temperature, where transport is defect-limited. 
For shallow donors, the impurities are fully ionized at higher temperatures, but only partially ionized or in their charge-neutral state at lower temperatures, posing additional challenges to our calculations. 
Figure~\ref{fig:mob-temper} shows the temperature-dependent electron mobility in P-doped Si; one set of results assumes fully-ionized impurities and the other takes into account partial impurity ionization as described in Supplemental Material~\cite{SM}.
%
%LATER:
%In both cases, the results are nearly identical for the full-ITA calculations shown in Fig.~\ref{fig:mob-temper}, which include $e$-ph and $e$-d scattering on the same footing, and for $e$-ph RTA plus $e$-d tr-RTA calculations (not shown).
%\\
%\indent
%
When accounting for partial ionization, our full ITA calculations with both $e$-ph and $e$-d scattering can accurately predict the experimental mobility~\cite{norton_impurity_1973} above $50$ K. However, scattering from the neutral P donor in our calculation is too weak to reproduce the measured residual mobility below $\sim$50~K.
This result suggests that the mobility below $50$~K may be limited by scattering mechanisms not included here. In particular, higher-order neutral-impurity scattering, which is usually neglected in the interpretation of transport experiments, has been shown to be important at low temperature~\cite{szmulowicz_calculation_1986}. 
Our results reinforce the hypothesis that low-temperature transport may be governed by higher-order $e$-d interactions with impurities in their charge-neutral state. A recent quantitative study using the T-matrix approach~\cite{kaasbjerg_atomistic_2020} concluded that higher-order $e$-d scattering rates for neutral defects can differ by orders of magnitude from the lowest-order $e$-d interactions employed here. More work is needed to include such higher-order effects in our transport calculations.
%In addition, one can consider the $e$-d RTs for neutral defects from other \textit{ab initio} frameworks such as Green's function method~\cite{ebert_calculating_2011}, and combine the RTs with the $e$-d RTs for charged defects and the $e$-ph RTs in the framework of BTE described in this work.***
\\
\indent
Our results highlight the need to carefully take into account partial dopant ionization.  
In Fig.~\ref{fig:mob-temper}, our full-ITA calculation assuming fully ionized donors gives electron mobilities well below the experimental values between 50$-$100~K, and saturate to a residual mobility an order of magnitude lower than experiment below $100$ K. The temperature trend of the mobility betweeen 50$-$100~K, where partial ionization is essential, is completely missed. 
%Although the computed carrier mobilities are consistent with the experimental data above $100$ K, they are underestimated below $100$ K, below which the free carriers from the shallow dopants are frozen out. 
%
%we thus consider the incompletely ionized case (please see Sec.~S4 and S6 of Ref.~\cite{SM} for details). 
%
%The refined carrier mobilities are improved and consistent with the experimental data above $40$ K.
%
%
%
%
%
%%%%%%%%%%
%% Discussion
%%%%%%%%%%
%
%
%
\\
\indent
The framework presented in this work lends itself to various applications beyond our proof-of-concept study of doped Si. Although we focused on shallow defects, our method can also be applied to deep-level defects, a topic of great relevance for halide perovskites and narrow-gap semiconductors. Both bulk and two-dimensional (2D) materials can be treated, provided the Coulomb potential in the 2D material is modified to account for the different dimensionality. Among other systems, van der Waals materials and heterostructures~\cite{novoselov_2d_2016}, as well as interfaces between bulk materials, would greatly benefit from detailed studies of how defects impact charge transport. Extensions to include higher-order neutral impurity scattering, for example with the T-matrix approach, will be considered in future work.
\newpage
%\\
%
\indent
% and better treatment of the screening 
%%%%%%%%%
% Conclusion
%%%%%%%%%
%%
In summary, we developed a rigorous and practical first-principles approach to compute $e$-d interactions due to ionized impurities or other charged defects. Our method can take into account the atomic structure of the defects, the spatially-varying Bloch wave functions, and an arbitrary, anisotropic and multi-valley band structure. 
This framework makes it possible to capture important defect physics in \textit{ab initio} calculations of electron dynamics. Our work enables studies of transport from low to high temperatures in doped semiconductors and oxides without any fitting or empirical parameters.
%
%This powerful tool for analyzing $e$-d interactions expands the scope of \textit{ab initio} studies of electron dynamics in real materials. 

%%%%%%%%%%%%%%%%%%%%%%%%%%%%%%%%%%%%%%%%%%%%%%%%%
% Acknowledgements
%%%%%%%%%%%%%%%%%%%%%%%%%%%%%%%%%%%%%%%%%%%%%%%%%
%
\begin{acknowledgments}
\vspace{-10pt}
This work was supported by the Air Force Office of Scientific Research through the Young Investigator Program Grant FA9550-18-1-0280. This research used resources of the National Energy Research Scientific Computing Center (NERSC), a U.S. Department of Energy Office of Science User Facility located at Lawrence Berkeley National Laboratory, operated under Contract No. DE-AC02-05CH11231. 
%J.-J. Z. was supported by the National Science Foundation under Grant No. ACI- 1642443, which provided for code development. 
%
%This research used resources of the National Energy Research Scientific Computing Center, a DOE Office of Science User Facility supported by the Office of Science of the U.S. Department of Energy under Contract No. DE-AC02-05CH11231.
%
I-T. L. thanks Dr.~Ivan Maliyov and Dr.~Cheng-Wei Lee for fruitful discussions.
\end{acknowledgments}

\bibliography{reference-cds}

\end{document}